# Cyber Resilience: by Design or by Intervention?

Alexander Kott[1], Maureen S. Golan[2,4], and Benjamin D. Trump[2,5*], Igor Linkov[2,3]

[1]US Army DEVCOM Army Research Laboratory, Adelphi, MD, USA

[2]US Engineer Research and Development Center, Boston, MA, USA

[3]Carnegie Mellon University, Pittsburgh, PA, USA

[4]Credere Associates, Concord, MA, USA

[5]University of Michigan, Ann Arbor, MI, USA

## Abstract

The term "cyber resilience by design" is growing in popularity. But what is the other resilience, not by design? In this article we explore differences and mutual reliance of resilience by design and resilience by intervention.

## Introduction

The term "cyber resilience by design" appears ever more frequently in the literature (Ahmadi-Assalemi et al. 2020; Bagchi et al. 2020; Gourisetti et al. 2019), and is a topic of this, third of a series of articles in this column on cyber resilience—an important attribute of system dependability. The previous two were (Linkov et. al. 2020) and (Kott and Linkov, 2021).

Before we move any further, let's orient ourselves with respect to our lexicon. Here, by cyber resilience we refer to the ability of the system to resist, minimize and mitigate a degradation caused by a successful cyber-attack on a system or network of computing and communicating devices (NRC 2012; Kott and Linkov, 2019). This is not the only definition of cyber resilience, but we see it as an important one, and will settle on it for the sake of this paper. But what does "by design" means?

Opinions differ. Some use the term "by design" when arguing that systems must be designed and implemented in a provable mission assurance fashion, with the system's intrinsic properties ensuring that a cyber-adversary is unable to cause a meaningful degradation (Jabbour 2019). Others envision autonomic

………………………..

resilience (Hariri et al. 2011) where cyber resilience is provided by reflexive, no-thinking-required actions of appropriate elements within the system. Yet others (Kott and Theron 2020) recommend that a system should include a built-in autonomous intelligent agent responsible for thinking and acting towards continuous observation, detection, minimization and remediation of a cyber degradation. In all cases, the qualifier "by design" indicates that the source of resilience is somehow inherent in the structure and operation of the system.

But what, then, is the other resilience, *not* by design? Clearly, there has to be another type of resilience, otherwise what's the purpose of the qualifier "by design"? Indeed, while mentioned less frequently, there exists an alternative form of resilience called "resilience by intervention" (Linkov et al. 2021). Unlike in the case of resilience by design (let's abbreviate it RBD), in resilience by intervention (RBI) the effects of an external actor are the necessary source of resilience.

To use a psychological analogy, most humans have some innate abilities to cope emotionally with an adverse effect, and to return to a pre-crisis state. Many cases, however, require an intervention by a professional, e.g., a therapist (Rynearson 2006). In case of a human, the distinction seems clear enough: resilience through self-coping mechanisms resident in the mind and body of a given individual reflects RBD, while engagement of an external actor – a therapist – is analogous to RBI. But does this distinction make sense in the world of cyber where systems are typically distributed, networked, and have fluid, permeable and uncertain boundaries? How would we even define what is an "external" actor?

## Three examples

To make our discussion more concrete, let's introduce three examples.

Example 1. Ms. Johnson owns a self-driving car. She purchased a service in which a third-party provider remotely, continuously monitors and assesses the cyber-health of her car, and issues alerts to Ms. Johnson if a compromise is detected. Then Ms. Johnson can take her car to any shop with cyber-repair capabilities, of her choice, and get the compromise eliminated.

Example 2. Ms. Johnson owns a self-driving car. The car includes a resident software agent responsible for continuously monitoring all events on the car's multiple computing devices and related internal and external networks; detecting and mitigating cyber compromises, and ensuring that the car can drive even when it is cyber-compromised.

Example 3. Ms. Johnson often gets around by calling a self-driving cab of Robo-Cab Inc. This company owns a fleet of self-driving cars, controlled and serviced from a company's depot. A depot-based system monitors each car, safely disables it when a compromise is suspected and sends a response team to the vehicle. In addition, every few hours, each car returns to depot for a thorough checkup and potential re-imaging.

Clearly, in each case, we find some provisions for cyber resilience. But is it RBI or RBD? Before we proceed with answering this question, let's offer two perspectives on what differentiates RBI and RBD.

## Integration and Authority

………………………

We find that differences between RBI and RBD of a given system are more clearly seen in light of two considerations. The first is the degree of integration of the resilience-producing elements with the rest of the system. RBD is likely to be found where strong links exist between the resilience elements of the system and those components of the system that are responsible for producing the primary functions of the system. On the other hand, if the links are established occasionally and selectively based on circumstances, we are probably looking at RBI.

(Parenthetically, we should mention that the term "system" in this paper refers to a socio-technical system consisting of technical artifacts as well as humans if and when humans are integral parts of the overall system; that is, the term encompasses the human, software, and hardware partitions of an information system. Therefore, when we refer to various capabilities of a system, those capabilities are allocated amongst the human, software, and hardware. Thus, when we refer to "intervention" and "external actor," the intervention may be performed by a strictly technical system (e.g., the external actor being an automated cyber-mitigation tool), or by humans, or by a socio-technical system that combines humans and technical means.)

The observation that tight integration is associated with RDB seems to be true regardless of the specific mechanisms by which resilience is established. In a case of a system designed with innate cyber-resilience properties, as advocated by (Jabbour 2019), the resilience mechanism is literally indistinguishable from the system's primary functional elements; implying RBD. In the case of a built-in autonomous intelligent agent responsible for cyber-resilience (Kott and Theron 2020), the agent is tightly integrated within the system, in an RBD fashion.

This is also true regardless of the nature of the system, be it, for example, technical or biological. For instance, human psychological coping mechanisms are inherent in a human, and as such illustrate RBD. A therapist on the other hand, is clearly not integral to the human patient, implying RBI.

In addition, the notion of integration does not necessarily imply spatial co-location. In a distributed system, various functions, including those that provide resilience, may reside, say, on different computers networked but separated by thousands of miles. This separation in itself tells us little about RBD or RBI. Instead, we should explore how closely and persistently these functions rely on each other.

The second perspective that helps differentiate between RBD and RBI is that of authority. When actions associated with resilience are taken, that is, actions of resisting, minimizing and mitigating a degradation caused by a successful cyber-attack, who has the authority to direct these actions? Who decides whether, when, where and how these actions are performed?

When these decision rights are internal to the system, it does not require the permission (or direction) of an external actor to change system properties or reorganize itself. This authority is important in order to seize quickly opportunities to minimize or to recover from disruption. A system with an internal decision-making authority is able to interpret new information within itself or its environment and adjust accordingly. The system benefits from avoiding potential inefficiencies or delays from waiting for an external actor to yield permission or direction for change.

Conversely, a system that lacks internal decision authority has to rely on an external intervention in order to recover or adapt system properties or organizing characteristics. Such systems may also have advantages -- it is possible that external intervention comes with greater decision-making abilities and

………………………..

may yield greater control over the intended future of a disrupted or adapted system. External and internal authorities may co-exist or conflict – we will return to this point shortly.

If the decision authority resides primarily outside of the system, we suspect RBI. If, on the other hand, the decisions are made primarily within the system (whether by humans or by a technical decision-making element), RBD is more likely.

Referring again to the emotional crisis example, in self-coping the decisions inevitably reside with the human experiencing the crisis. However, when a therapist intervenes, the patient may express his or her wishes, but the therapist reserves the ultimate authority on how and whether to provide a treatment. This is characteristic of RBI.

## Back to Three Examples

Now we are ready to return to Ms. Johnson and her means of transportation. Let us use the two perspectives we just discussed – integration and authority – to explore whether our three examples represent RBD or RBI.

In the first example, the integration links between Ms. Johnson and her car (together they are "the system") on the one hand, and the cyber-services providers are rather weak. If Ms. Johnson is unhappy with the service – or finds a better price – she will readily switch to another provider. She probably does not adapt her routine or her car to the needs of the provider; she would prefer to keep her options of choosing among multiple providers. Likewise, the providers are unlikely to customize their service to meet Ms. Johnson's unique requirements. They might try to entice her into a long-term contract, but otherwise Ms. Johnson is just one of many customers. Their mutual reliance is minimal.

And what about authority? Ms. Johnson may express her wishes, but ultimately the service providers will decide how they provide the service, when (e.g., "sorry, this week we are too busy") and which specific processes and procedures they will use. Ms. Johnson and her car do not have much authority over these decisions. Let's conclude: integration is minimal and authority is external to the system. This is RBI.

In the second example, the resilience-producing component – a software agent responsible for monitoring and mitigation of compromises – is fully integrated into the overall system. It is customized or custom-designed for the given car, and for the pattern of operations characteristic of consumers like Ms. Johnson. It is dedicated entirely to the car and to Ms. Johnson, and does not have much purpose outside of that system. The integration is tight.

The authority for making decisions about how to monitor the cyber-state of the car, how to detect and analyze the compromise, and what mitigating actions to take – all resides with the agent. Granted, the agent will attempt to ask for Ms. Johnson's approval in appropriate situations, but in most cases the agent will have to act autonomously because Ms. Johnson is unavailable or unable to make a relevant decision. Here we have tight integration of resilience-producing elements into the system, and decision-making authority residing internally in the system. This is RBD.

However, although rare, some impacts to this cyber system may be too large, outside the scope of the software agent's mitigation or repair capabilities, or cause myriad cascading impacts that require Ms. Johnson to seek specific service capabilities elsewhere. In these circumstances, Ms. Johnson does not have any prior commitments or contracts to any service providers, and the integration is loose. And

……………………….

although the monitoring service has authority over the vehicle under situations it is capable of mitigating, should Ms. Johnson require to seek services elsewhere during exceptional events, the single-job provider will have the onus of recovery and adaptation placed on them. They will likely collaborate or seek system history from Ms. Johnson's cyber monitor, but a majority of the authority is external to the system. This scenario therefore is RBI.

The third example differs significantly from the previous two. Importantly, the system under consideration is different. It consists of the fleet of cars and the depot. Ms. Johnson is not a part of the system, she is an external consumer of the services produced by the system.

All activities related to cyber resilience are integrated within the depot-cars system. Note that the system is highly distributed and dynamic: the depot itself may consist of multiple physical facilities, and its cyber-related systems may reside on multiple computers, anywhere in the world. The cars are traveling over a wide geographic area, constantly changing their locations. Nevertheless, all cyber-resilience components and their activities are integrated within this distributed system. The depot's resilience-producing components (including human specialists) are designed to support the fleet of cars, and cars cannot operate (at least for this business model) without the depot support.

Regarding the authority for making cyber resilience related decisions, they also reside substantially within the depot-cars system (including its human employees). The decisions on how to resist, minimize and mitigate a degradation caused by a successful cyber-attack are made by automated or human elements within the overall system, even if the decision may be produced in a distributed fashion between multiple elements of the system. Thus, we have tight integration and internal authority, implying RBD.

Table 1. Comparison of illustrative examples and their implications for cyber resilience implementation.

|  | **Example 1** | **Example 2** | **Example 3** |
|---|---|---|---|
| **Overview** | Ms. Johnson owns self-driving car and has a contract with a third-party continuous monitoring provider | Ms. Johnson owns self-driving car with a resident software agent for continuous monitoring. | Ms. Johnson uses a self-driving car service with a depot-based continuous monitoring system |
| **System** | Ms. Johnson and her car | Ms. Johnson and her car with monitoring | Robo-Cab, Inc. (fleet of cars, depot, monitoring) |
| **Goal** | Ms. J & car: maintain safe mobility and quick recovery from cyberthreats | Ms. J & car: maintain safe mobility and seamless recovery from cyberthreat | Robo-Cab: maintain safe mobility for customers and seamless recovery from cyberthreat |
| **Environment** | Third-party monitoring, alerts, and choice of cyber-repair shop | Any self-driving car shop/garage | Ms. Johnson and her mobility needs |
| **Integration** | Loose: Weak link between Ms. J & repair process | Tight: Autonomous response by software agent (represent Ms. J) & repair process  Loose: Certain disruptions require one-time contracts with specialty shops | Tight: Fleet, depot, and support uniquely designed for the distributed system |
| **Authority** | External: Service providers decide path forward & responsible for repair outcome | Internal: Resident software agent holds responsibility for repair  External: Specialty shop responsible for repair | Internal: Robo-Cab's cyber-physical-human team hold responsibility for repair |
| **Resilience Type** | Resilience-by-intervention | Resilience-by-design & Resilience-by-intervention | Resilience-by-design |
| **Advantage** | Response is more threat & impact dependent; resources used as needed | Response is more immediate & consequences may be less severe | Response is more immediate & consequences may be less severe |

……………………..

| **Disadvantage** | May be too slow of a response | Cyberthreat may adapt to & overcome | Cyberthreat may adapt to & overcome |

## Design and Intervention should be Complementary

Needless to say, RBD and RBI, although different, are not incompatible. Both approaches may co-exists. A given system may have provisions for both RBD and RBI, although this co-existence should be carefully orchestrated. In particular, a clear protocol should be established for the handover of responsibilities between RBD mechanisms to RBI mechanisms, and back. If both RBI and RBD mechanisms need to operate simultaneously, a coordination protocol should ensure that their respective actions do not produce undesirable interference. Interactions of RBD and RBI should be specific to system and situation.

Furthermore, both RBD and RBI have their disadvantages. For example, relying strictly on RBI, as in Ms. Johnson's first example, may be risky. If RBI-based mitigation of a cyber compromise cannot be provided rapidly, she might find herself speeding on a highway in a dangerously uncontrollable car. For such emergency situations, it would be far safer to provide her car with an on-board RBD mechanism, like in example 2, even if it is less capable then comprehensive cyber-resilience service by intervention. As we discussed elsewhere (Kott and Theron 2020), Ms. Johnson does live in an environment where fast, brutal cyberattacks exist, and demand extremely fast response. Criminals or irresponsible pranksters might take control of Ms. Johnson's car, and consequences can be tragic (Ring 2015). That's why her car would benefit from having an onboard intelligent autonomous agent capable of taking the necessary resilience actions, that is, an RBD component.

Similarly, reliance only on RBD comes with its own risks. For example, if a cyber attacker is able to overcome the capabilities – inevitably limited – of the RBD mechanisms, external intervention is a necessity. Such an intervention may come from multiple external actors, depending on the specific challenge of a cyber-compromise. Even if most of the external actors find themselves unable to handle the complex compromise, chances are relatively high that in a multiplicity of external actor, at least one is found who is able to deal with the challenge.

Provisions for adding RBI to RBD, when a situation dictates, are necessary because sooner or later such a situation would arise. This conclusion is no by no means unique to cyber resilience. For example, in case of supply-chain, engineering and environmental resilience, the use of resilience analytics facilitates the implementation of corrective actions from within the system and/or external to the system for a unified approach that maintains necessary critical function despite inevitable disruption (Golan et al. 2021; Linkov et al. 2021; Simchi-Levi and Simchi-Levi 2020; Trump et al. 2020).

To conclude, the terms "cyber resilience" and "resilience by design" are gaining popularity. Occasionally they seem to be misused as synonyms or as a "cooler" substitute for tired terms like cyber security, with little concrete meaning behind them. This should not be the case. Resilience by design is a particular type of cyber resilience (not to be confused with cyber security), distinct from resilience by intervention and perhaps other types, yet to be identified. Both types require clear definition and differentiation. To distinguish RBD and RBI, it is useful to examine how tightly the resilience-producing elements are

………………………..

integrated into the overall system, and whether the authority for making resilience-related decisions resides within or outside the system. Neither RBD nor RBI are a panacea, and careful integration of the two is likely to produce superior resilience.

Table 2. Comparison of risk management approaches (i.e., cybersecurity), resilience-by-design, and resilience-by-intervention for cyber systems.

|  | **Risk management** | **Resilience-by-design** | **Resilience-by-intervention** |
|---|---|---|---|
| **Objective** | Harden individual components | Design components to be self-reorganizable | Rectify disruption to components and stimulate recovery by external actors |
| **Capability** | Predictable disruptions, acting primarily from outside the system components | Either known/predictable or unknown disruptions, acting at a component or system level | Failure in context of societal needs, may be constellation of networks across systems |
| **Consequence** | Vulnerable nodes and/or links fail as result of threat | Degradation of critical functions in time and capacity to achieve system's function | Degradation of critical societal function due to cascading failure in interconnected networks. |
| **Actor** | Either internal or external to the system | Internal to the system | External to the system |
| **Corrective Action** | Either loosely or tightly integrated with the system | Tightly integrated with the system | Loosely integrated with the system |
| **Stages/Analytics** | Prepare and absorb (risk is product of threat, vulnerability and consequences and is time independent) | Recover, and adapt (explicitly modeled as time to recover system function and the ability to change system configuration in response to threats) | Prepare, absorb, recover, and adapt (explicitly modeled as ability to recover and secure critical societal function and needs through constellation of relevant systems) |

# References

Ahmadi-Assalemi, G., Al-Khateeb, H., Epiphaniou, G., Maple, C. (2020). Cyber Resilience and Incident Response in Smart Cities: A Systematic Literature Review. Smart Cities, 3(3): 894-927. https://doi.org/10.3390/smartcities3030046

Bagchi, S., Aggarwal, V., Chaterji, S., Douglis, F., Gamal, A. E., et al. (2020). Vision Paper: Grand Challenges in Resilience: Autonomous System Resilience through Design and Runtime Measures. IEEE Open Journal of the Computer Society, 2020(1): 155-172. https://doi.org/10.1109/OJCS.2020.3006807

Gourisetti, S. N. G., Scott, M., Michael, M., et al. (2019). Secure Design and Development Cybersecurity Capability Maturity Model (SD2-C2M2): Next-Generation Cyber Resilience by Design. Proceedings of the Northwest Cybersecurity Symposium, 2019. https://doi.org/10.1145/3332448.3332461

Hariri, S., Eltoweissy, M., Al-Nashif, Y. (2011). BioRAC: biologically inspired resilient autonomic cloud. Proceedings of the Seventh Annual Workshop on Cyber Security and Information Intelligence Research, 80: 1-1. https://doi.org/10.1145/2179298.2179389

……………………..

Jabbour, K. (2019). The Post-GIG Era. The Cyber Defense Review, 4(2): 117-128. https://cyberdefensereview.army.mil/CDR-Content/Articles/Article-View/Article/2017729/the-post-gig-era-from-network-security-to-mission-assurance/

Golan, M. S., Trum, B. D., Cegan J. & Linkov, I. (2021). The Vaccine Supply Chain: A Call for Resilience Analytics to Support COVID-19 Vaccine Production and Distribution. COVID: Risk and Resilience. Springer International Publishing, New York, NY. https://arxiv.org/abs/2011.14231

Kott, A., Theron, P. (2020). Doers, not watchers: Intelligent autonomous agents are a path to cyber resilience. IEEE Security & Privacy, 18(3): 62-66. http://doi.org/10.1109/MSEC.2020.2983714

Kott, A., Linkov. I., eds. (2019). Cyber Resilience of Systems and Networks. Switzerland: Springer. https://doi.org/10.1007/978-3-319-77492-3

Kott, A., Linkov, I., (2021) To Improve Cyber Resilience, Measure It. Computer, 54(2), 80-85. https://doi.ieeecomputersociety.org/10.1109/MC.2020.3038411

National Research Council (NRC). (2012). Disaster Resilience: A National Imperative. Washington, DC: The National Academies Press. https://doi.org/10.17226/13457

Linkov, I., Trump, B. D., Golan, M., Keisler, J. M. (2021). "Enhancing Resilience in Post-COVID Societies: By Design or By Intervention?" Environmental Science & Technology, 55(8):4202-4204. https://doi.org/10.1021/acs.est.1c00444

Linkov, I., Galaitsi, S., Trump, B. D., Keisler, J. M., & Kott, A. (2020). Cybertrust: From Explainable to Actionable and Interpretable Artificial Intelligence. *Computer*, *53*(9), 91-96.

Ring, T. (2015). Connected cars – the next target for hackers. Network Security, 2015(11): 11-16. https://doi.org/10.1016/S1353-4858(15)30100-8

Rynearson, E. K., ed. Violent death: Resilience and intervention beyond the crisis. New York, NY: Routledge, 2006. ISBN: 0-415-95323-5.

Simchi-Levi, D., and Simchi-Levi, E. (2020). We Need a Stress Test for Critical Supply Chains. Harvard Business Review. 28 April 2020. https://hbr-org.cdn.ampproject.org/c/s/hbr.org/amp/2020/04/we-need-a-stress-test-for-critical-supply-chains

Trump, B. D., Linkov, T., Hynes, W. (2020). Combine Resilience and Efficiency in Post-COVID Societies. Nature, 588(220). https://doi.org/10.1038/d41586-020-03482-z

………………………..

## Author Bios

ALEXANDER KOTT is the Chief Scientist with the U.S. Army DEVCOM Army Research Laboratory, Adelphi, Maryland. Contact him at alexander.kott1.civ@mail.mil.

BENJAMIN D. TRUMP is a research social scientist with the U.S. Army Corps of Engineers' Engineer Research and Development Center, Concord, Massachusetts. Contact him at benjamin.d.trump@usace.army.mil.

MAUREEN S. GOLAN is a research engineer with the U.S. Army Corps of Engineers' Engineer Research and Development Center, Concord, Massachusetts. Contact her at maureengolan@gmail.com.

IGOR LINKOV is a Senior Science and Technology Manager with the U.S. Army Corps of Engineers' Engineer Research and Development Center, Concord, Massachusetts. Contact him at Igor.linkov@usace.army.mil.

DISCLAIMER